\documentclass[useAMS,usenatbib]{mn2e}
\usepackage[dvips]{epsfig}

\begin{document}
\label{firstpage}

\title{Persistent Patterns in Accretion Disks}
\author[M.~A.~Amin and A.~V.~Frolov]{%
Mustafa A. Amin and Andrei V. Frolov%
\thanks{E-mail: \texttt{mamin@stanford.edu}, \texttt{afrolov@stanford.edu}}\\%
KIPAC, Stanford University, Stanford, CA, 94305-4060}

\date{24 March 2006} \volume{(to be submitted)} \pagerange{\pageref{firstpage}--\pageref{lastpage}} \pubyear{2006}

\maketitle

\begin{abstract}
We present a set of new characteristic frequencies associated with
accretion disks around compact objects. These frequencies arise from
persistent rotating patterns in the disk that are finite in radial
extent and driven purely by the gravity of the central body. Their
existence depends on general relativistic corrections to orbital motion
and, if observed, could be used to probe the strong gravity region
around a black hole. We also discuss a possible connection to the puzzle
of quasi-periodic oscillations.
\end{abstract}


\section{Introduction}

Timing observations of accreting X-ray binary systems have revealed
luminosity modulation at a number of characteristic frequencies.
Phenomenology of these quasi-periodic oscillations (QPOs) is quite rich.
For a detailed review, see \citet{2003astro.ph..6213M} and references
therein. Some of the features are rather puzzling, such as stability of
high frequency QPOs in black hole binaries and that in some
systems they appear in pairs at 3:2 frequency ratio.

QPOs in black hole systems are thought to arise from physical processes
in accretion disks. Depending on where the oscillations reside, one can
roughly divide models for QPOs in accretion disks into two classes:
local and global. Local models tie down the oscillation frequency to a
particular place in the disk (like an edge or a hot spot). In this case,
the question of what determines that place has to be answered. One line
of argument is that the location of the hot spot is determined by a
resonance
\citep{2001A&A...374L..19A,2005A&A...436....1T,2005AN....326..820K,2005AN....326..782A}.
This model has an attractive feature that the observed 3:2 frequency
ratio can be explained by non-linear mode locking. A hot spot can give
rise to luminosity variation, for example, due to Doppler beaming
\citep{2004ApJ...606.1098S,2005ApJ...621..940S}. However, a potential
difficulty is to have a hot spot which is sufficiently bright. Achieving
sufficient luminosity variation seems less problematic in global models,
in which modes occupy a larger region of the disk. Linear perturbation
analysis of the accretion disk in diskoseismology approach
\citep{2001ApJ...559L..25W,1997ApJ...476..589P,2001ApJ...548..335S,2002ApJ...567.1043O}
naturally solves the issue of spatial and frequency localization of
modes. The 3:2 frequency ratio would be accidental for two fundamental
diskoseismic modes, but it could arise from higher azimuthal $g$-modes
which are nearly harmonic.

In this paper, we describe a set of new characteristic frequencies which
might be present in accretion disks around compact objects. To the best
of our knowledge, they have remained unnoticed in the literature. These
frequencies arise from rotating patterns in the disk which are
quasi-stationary, finite in radial extent, and driven purely by gravity
of the central body. We neglect self-gravity and the hydrodynamics of the
accreting matter. The main idea is similar to the notion of density
waves that give rise to the spiral structure in galaxies
\citep{1963StoAN...5.....L,1964ApJ...140..646L,1966PNAS...55..229L},
although these patterns depend on general relativity rather than a
distributed matter source for their existence.

While it is tempting to identify the frequencies of these patterns with
the source of QPOs, we cannot claim to have a complete model. The issues
of how they are excited, how they translate to X-ray luminosity
variation, and effects of pressure and viscosity need to be investigated
in more detail. We will return to these points with some plausibility
arguments in Section~\ref{sec:discussion}.

\section{Accretion Disk Kinematics}

A test particle in a circular equatorial orbit around a Kerr black hole
has an orbital frequency \citep{1972ApJ...178..347B}
\begin{equation}
\Omega = \frac{1}{r^{3/2} + a}
%       = \frac{32.3{\rm kHz}}{r^{3/2} + a}\, \frac{M_{\sun}}{M}
\end{equation}
with respect to Boyer-Lindquist time $t$, where $r$ is the orbit radius
and $a$ is the dimensionless black hole spin parameter ($a=cJ/GM^2$). We
work in dimensionless units scaled by the black hole mass $M$ (i.e.,
distances measured in units of $GM/c^2$, times measured in units of
$GM/c^3$, etc.), and will further set $G=c=1$. Here and later we will
assume that particles co-rotate with the black hole.

\begin{figure}
  \epsfig{file=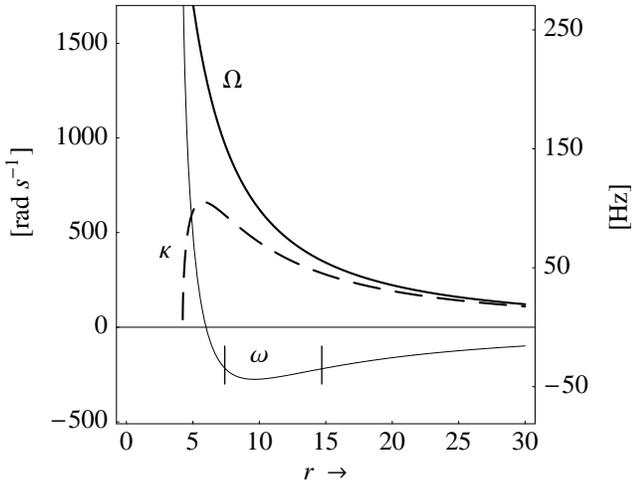, width=20pc}
  \caption{
    Orbital frequency $\Omega$, radial epicyclic frequency $\kappa$, and
    precession frequency $\omega = \Omega-2\kappa$ of a 2:1 orbit in an
    accretion disk around a Kerr black hole with $M=10\,M_{\sun}$ and
    $a=1/2$. Precession frequency exhibits a shallow negative minimum at
    $r_* \approx 9.64$. The radial extent of the rotating pattern with
    20\% deviation in frequency is marked by vertical bars.
  }
  \label{fig:freq}
\end{figure}

If perturbed from the circular orbit, the particle will undergo radial
and perpendicular oscillations with epicyclic frequencies $\kappa$ and
$\Omega_\perp$ respectively \citep{1987PASJ...39..457O}
\begin{equation}
\kappa^2 = \Omega^2 \left(1 - \frac{6}{r} + \frac{8 a}{r^{3/2}} - 3\, \frac{a^2}{r^2}\right)
\end{equation}
and
\begin{equation}
\Omega_\perp^2 = \Omega^2 \left(1 - \frac{4 a}{r^{3/2}} + 3\, \frac{a^2}{r^2}\right).
\end{equation}
The factors multiplying $\Omega$ on the right-hand sides of these
expressions are general relativistic corrections. They are absent in
Keplerian mechanics, where both epicyclic and orbital frequencies are
all the same ($\kappa=\Omega_\perp=\Omega$). The radial dependences of
orbital frequency $\Omega$ and radial epicyclic frequency $\kappa$ for a
typical rotating black hole are illustrated in Figure~\ref{fig:freq}.
Circular orbits close to a black hole are unstable; the innermost stable
circular orbit (ISCO) is located where $\kappa^2$ vanishes.

If orbital and epicyclic frequencies are the same, as they are for a
Keplerian potential, the orbits are closed. However, if the potential
deviates from $1/r$ (either because of general relativity corrections,
as in our case, or due to a distributed matter source, as happens in
galaxies), the two frequencies will in general be different, and the
orbits will precess. The condition for an orbit to close in a frame
rotating with frequency $\omega$ is for the orbital and epicyclic
frequencies to be commensurate, $m(\Omega-\omega) = n\kappa$, which
gives the precession frequency
\begin{equation}
\omega = \Omega - \frac{n}{m}\, \kappa.
\end{equation}
The integers $n$ and $m$ determine the shape of the precessing orbit,
and from here on, we will use the abbreviation $n$:$m$ to refer to their
values. Figure~\ref{fig:orbits} shows the shape of 1:2 and 2:1 orbits,
which are representative of the deformed and the self-intersecting orbit
classes.

In general, precession frequency $\omega$ depends on $r$, and any
pattern initially present will shear away as the disk rotates. However,
if $\omega$ is approximately constant over some portion of the disk,
collective orbit precession can lead to a nearly rigid pattern rotation.
One example of this is the spiral structure in galaxies caused by the
1:2 mode \citep{1963StoAN...5.....L,1964ApJ...140..646L,1966PNAS...55..229L}.
The 1:1 mode in Keplerian disks gives rise to a static one-armed spiral
pattern \citep{1983PASJ...35..249K}, which is seen in numerical
simulations as well \citep{2005MNRAS.360L..15H}. An attempt has been
made to trap the 1:1 mode in the region of the strong gravity
\citep{1990PASJ...42...99K}, but trapping depends strongly on the
pressure distribution within the disk \citep{1991PASJ...43...95K}.

\section{Persistent Patterns in the disk}

Our key observation is that for $n>m$, the precession frequency $\omega(r)$
develops a very shallow minimum at a radius $r=r_*$, as illustrated in
Figure~\ref{fig:freq} for the 2:1 orbit. Collective excitation of particles
on orbits precessing at the same rate would lead to a pattern occupying
a sizable portion of the disk around $r_*$ and rotating with little
shear at a frequency $\omega_p=\omega(r_*)$. Somewhat unusual are the
facts that the pattern is counter-rotating and that the orbit closes in
several rotations rather than a single one.

\begin{figure}
  \begin{center}
  \begin{tabular}{c@{\hspace{3pc}}c}
    {\LARGE\sf 1:2} &
    {\LARGE\sf 2:1} \medskip\\
  \epsfig{file=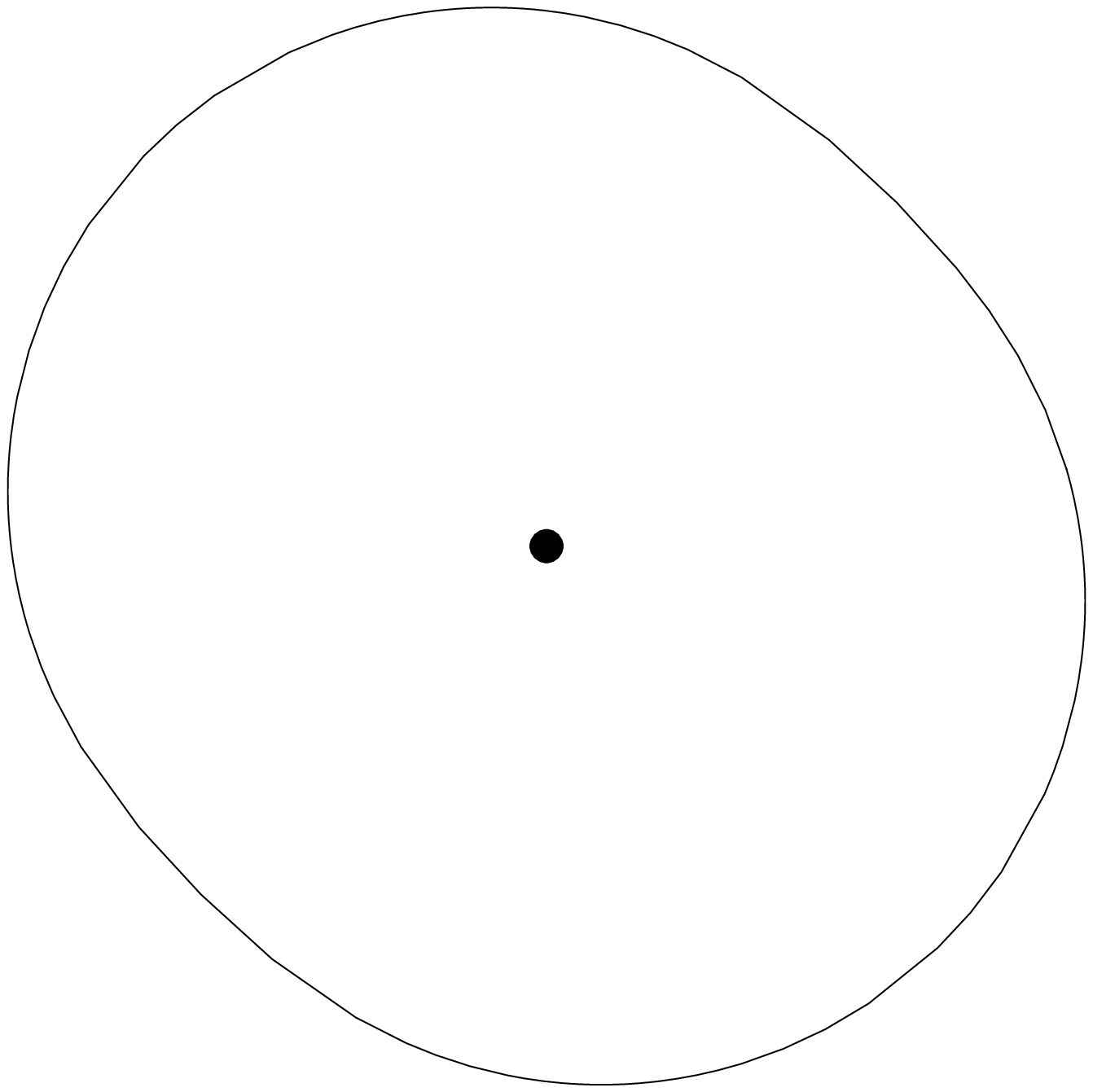, width=7pc} &
  \epsfig{file=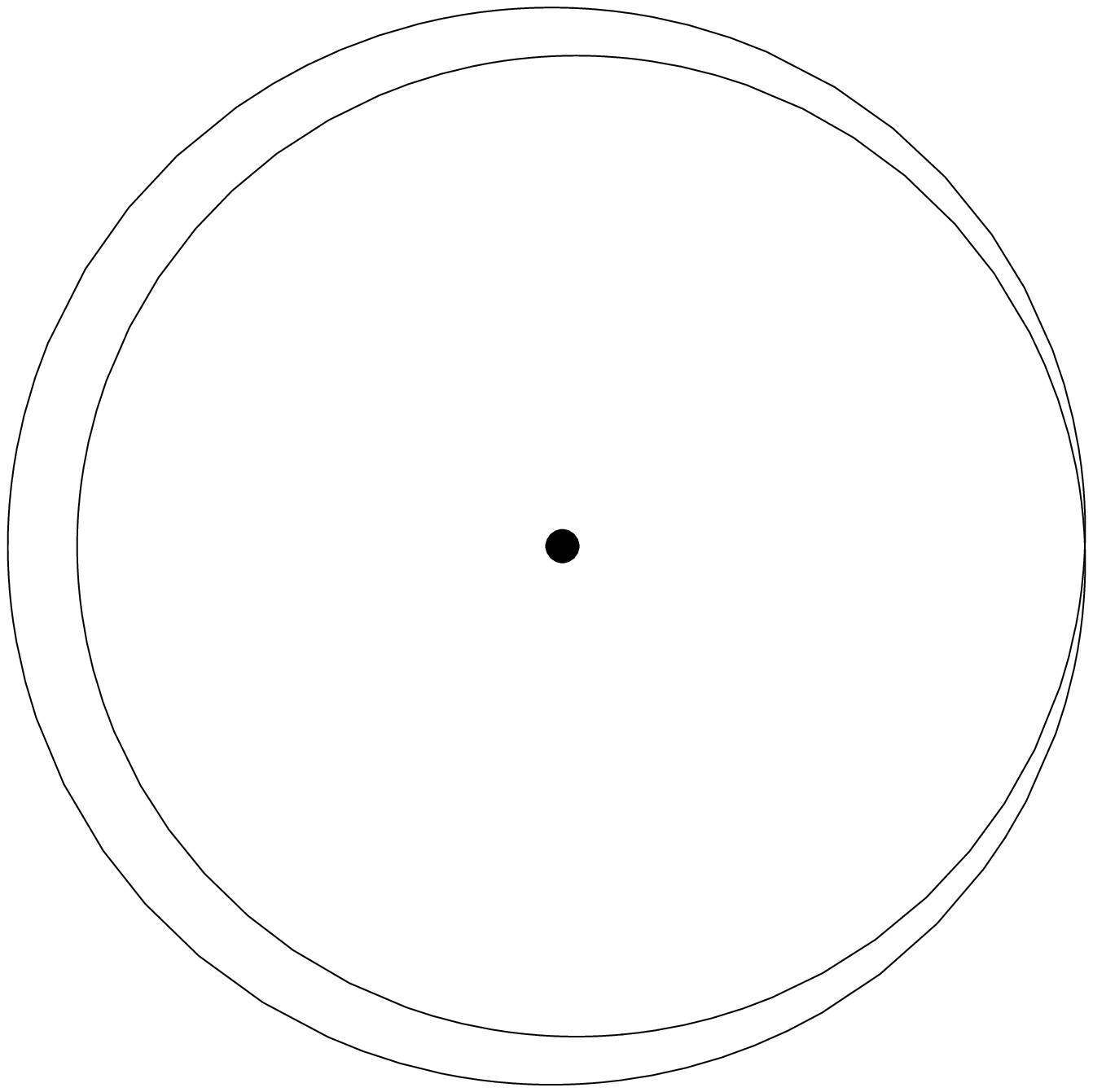, width=7pc}
  \end{tabular}
  \end{center}
  \caption{
    Sample closed orbits. 1:2 orbit (left) corresponds to 2 epicycles
    per 1 rotation around a central body, while 2:1 orbit (right)
    corresponds to 1 epicycle per 2 rotations.
  }
  \label{fig:orbits}
  \begin{center}
  \begin{tabular}{@{\hspace{-0.5pc}}c@{\hspace{0pc}}c}
    {\LARGE\sf \hspace{1pc}1:2} &
    {\LARGE\sf \hspace{1pc}2:1} \smallskip\\
    \epsfig{file=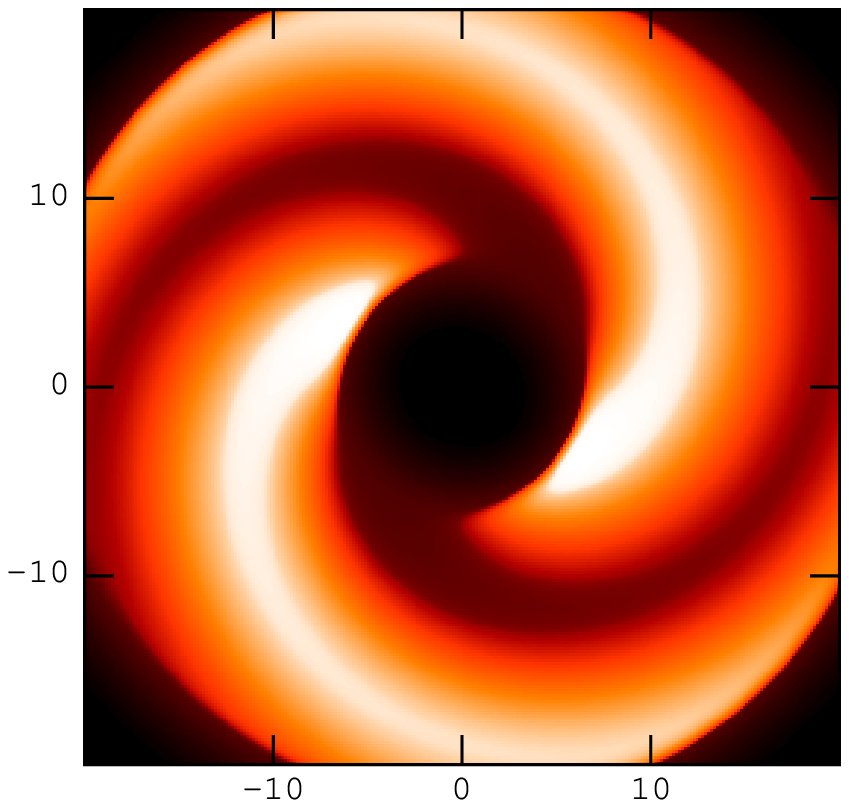, width=10pc} &
    \epsfig{file=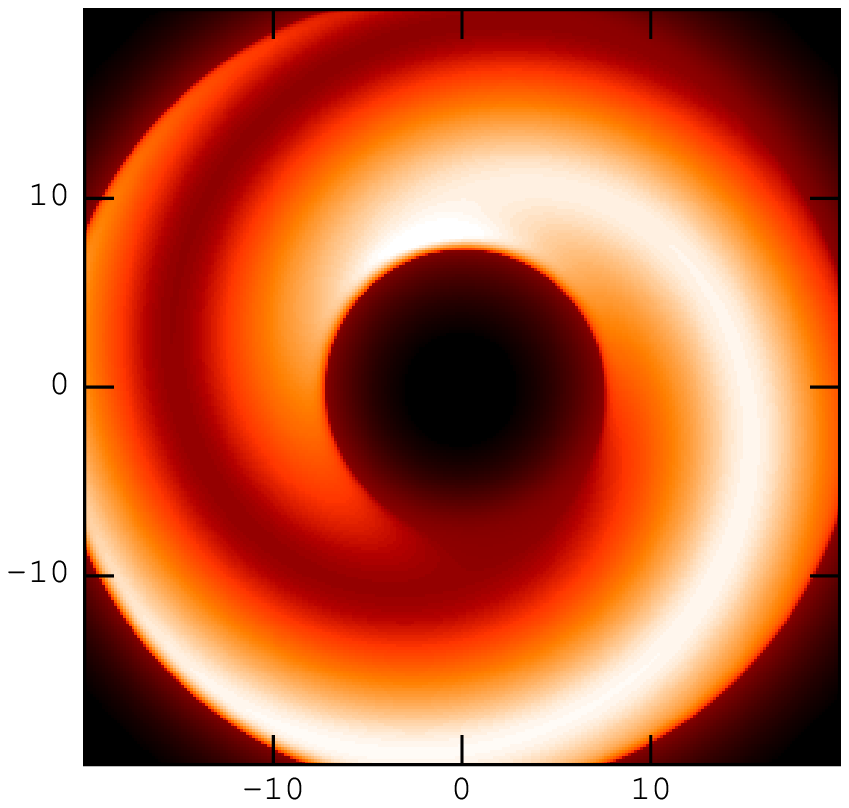, width=10pc} \\
    \epsfig{file=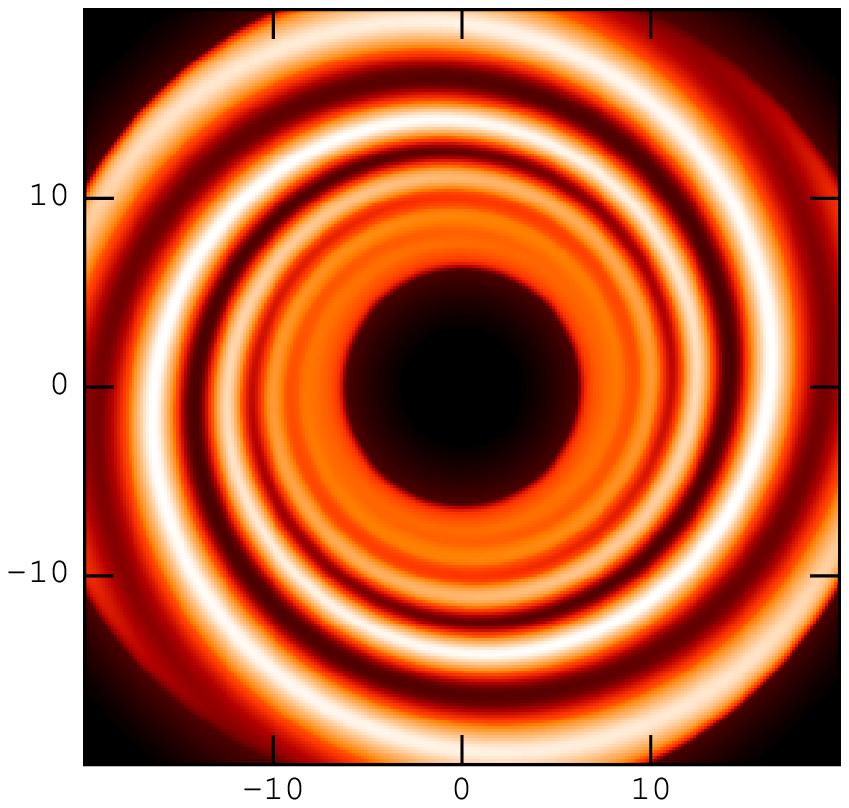, width=10pc} &
    \epsfig{file=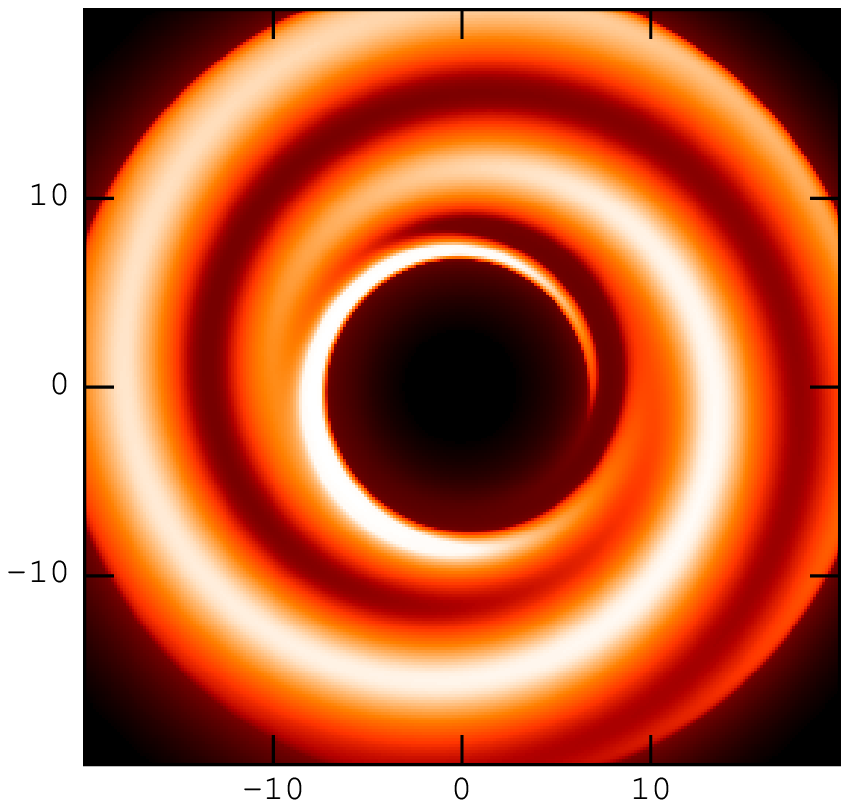, width=10pc} \\
    \epsfig{file=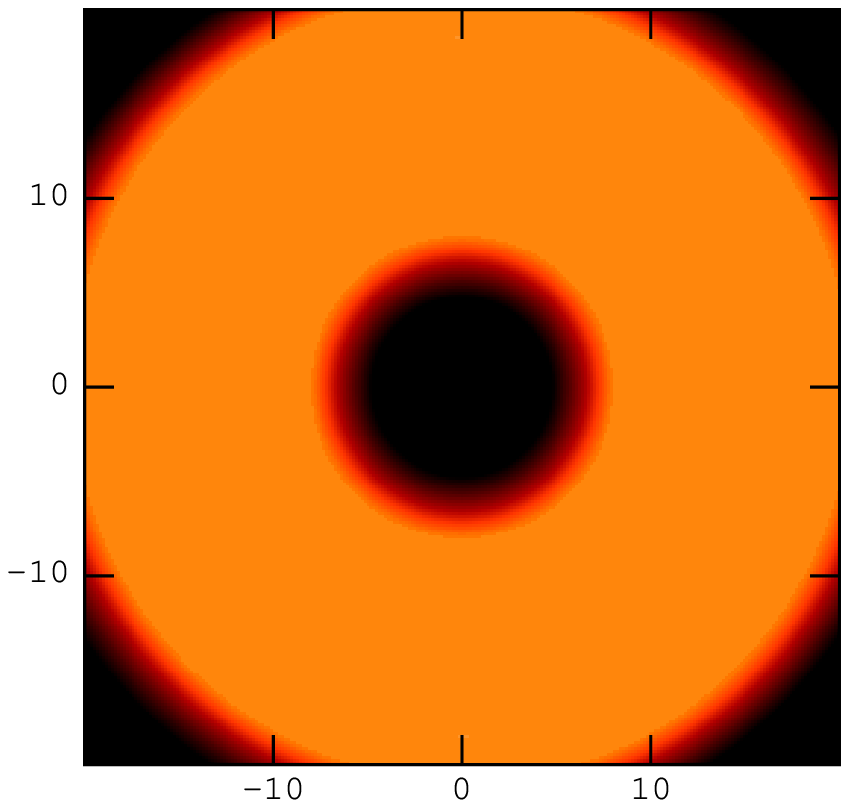, width=10pc} &
    \epsfig{file=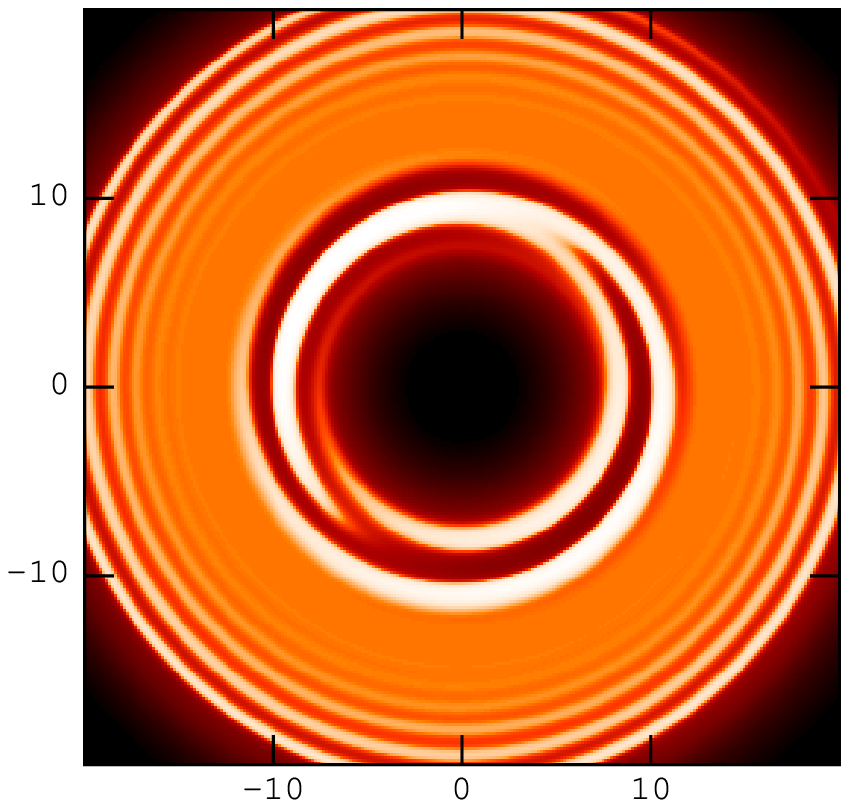, width=10pc} \\
  \end{tabular}
  \end{center}
  \caption{
    Evolution of density patterns obtained by distributing particles on
    stacked 1:2 (left) and 2:1 (right) orbits. Top row shows initial
    configuration, middle row - after one period of rotation of 2:1
    pattern elapsed, bottom row - after twenty periods. Density contrast
    has been enhanced by histogram equalization.
  }
  \label{fig:patterns}
\end{figure}

In a real accretion disk, the collective particle motion would have to
be excited by some dynamical mechanism. It might be complicated and
could require numerical simulations of the disk to fully understand the
driving process. In the present paper, we will be content with studying
the kinematics only. We will set off a collective mode by selecting
appropriate initial conditions (as described below) and follow the
pattern evolution by tracing the motion of individual particles making
up the disk. The purpose is to find out if there is a long-lived pattern
that survives the differential rotation.

If perturbed from a circular orbit at $r=r_0$ by a small displacement
$\varepsilon$ in the radial direction, the trajectory of a test particle
(to first order in $\varepsilon$) is
\begin{equation}
r(t) = r_0 + \varepsilon \sin (\kappa t + \chi),
\end{equation}
\begin{equation}
\phi(t) = \varphi + \Omega t + \frac{2\Omega}{\kappa r}\, \Upsilon\varepsilon \cos (\kappa t + \chi),
\end{equation}
where $\varphi$ and $\chi$ are initial orbital and epicyclic phases, and
\begin{equation}
\Upsilon = \frac{1 - \frac{3}{r} + \frac{2a}{r^{3/2}}}{1 - \frac{2}{r} + \frac{a^2}{r^2}}\, \left(r^{3/2} \Omega\right)
\end{equation}
is a relativistic correction factor (which, however, changes little in
the region of the disk we are interested in). We populate the disk by
spreading $N$ particles uniformly on a $n$:$m$ orbit, with initial
phases of a $k^{\mbox{\scriptsize th}}$ particle
\begin{equation}
\varphi_k = \frac{2\pi n}{N}\, k,\hspace{1em}
\chi_k = \frac{2\pi m}{N}\, k,
\end{equation}
while stacking the orbits in a radial direction at an angle $\alpha$ by
giving the orbit located at $r_j$ a phase offset
\begin{equation}
\varphi_{jk} = \varphi_k + \alpha r_j.
\end{equation}
This particle distribution leads to a spiral structure in the disk.
Figure~\ref{fig:patterns} shows the surface density contrast (smoothed
with a Gaussian kernel) for patterns obtained by distributing particles
on stacked 1:2 (left) and 2:1 (right) orbits. The three rows of
Figure~\ref{fig:patterns} show a time-lapse sequence of pattern
evolution. The top row shows the initial conditions, and the second and
the third rows show patterns at $t = T_{2:1}$ and $t = 20\,T_{2:1}$
correspondingly. $T_{2:1} = 2\pi/\omega_{2:1}$ denotes a period of
rotation of the 2:1 pattern.

The frequency of the 1:2 orbit precession depends monotonically on the
radius, so one expects differential rotation to destroy the pattern.
Indeed, at $t=T_{2:1}$, the spiral is seen to wind up, and by
$t=20\,T_{2:1}$, it is wound up so tightly that the smoothing removes
all traces of structure. The evolution of the 2:1 pattern is markedly
different. Signs of shear are clearly seen after a single rotation.
However, even after twenty rotations, there is still a pattern present
around $r_*\approx9.6$ (which is exactly where the minimum of
$\omega_{2:1}$ occurs). As this time span corresponds to almost 50
orbital rotations at $r_*$, the pattern is remarkably persistent.

\section{Discussion}\label{sec:discussion}

\begin{figure}
  \epsfig{file=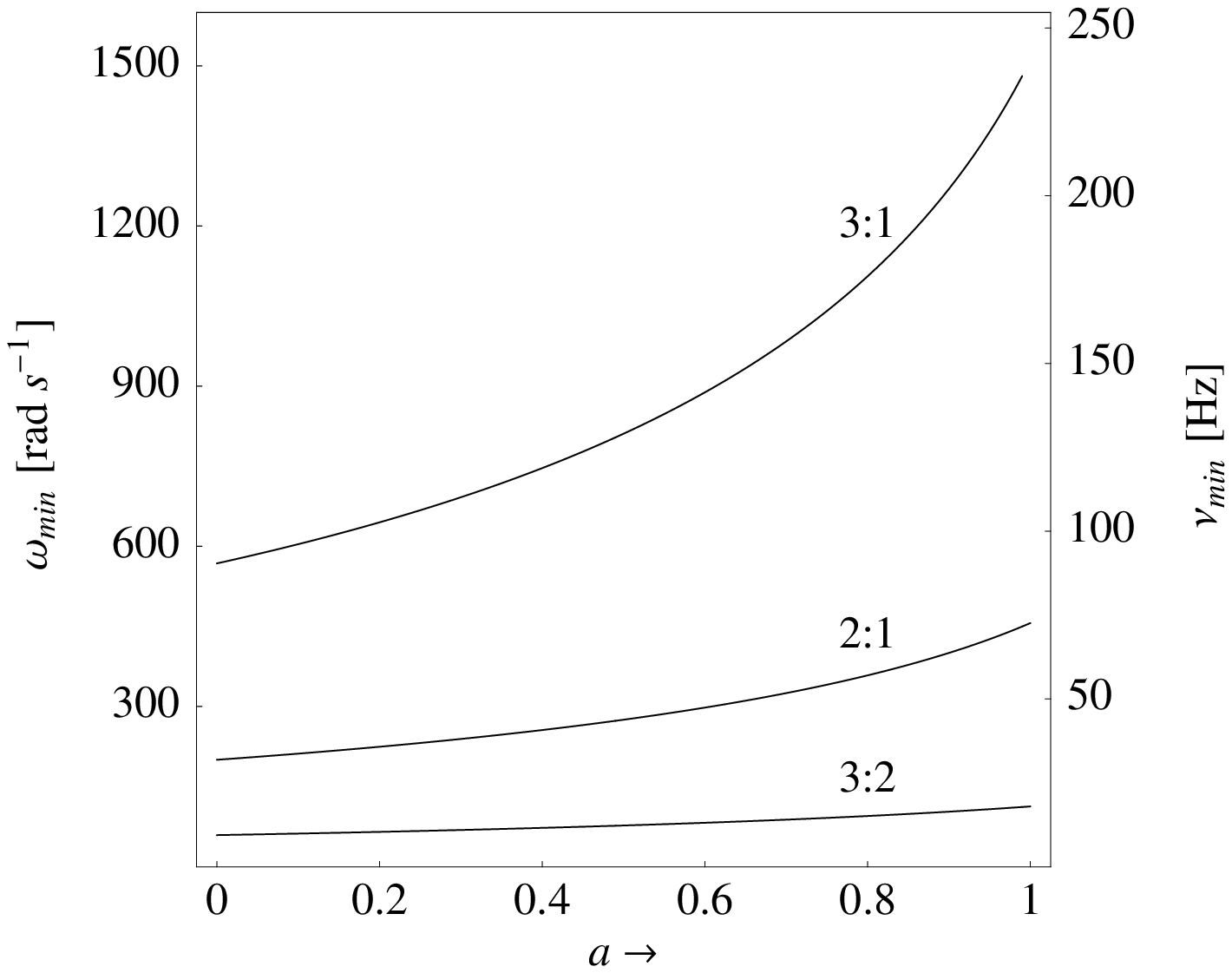, width=20pc}
  \caption{
    Persistent pattern frequencies of three lowest-order radial modes
    (2:1, 3:1, and 3:2) for a $10M_{\sun}$-mass black hole as a function
    of black hole spin parameter $a$.}
  \label{fig:wa:rad}
  \medskip
  \epsfig{file=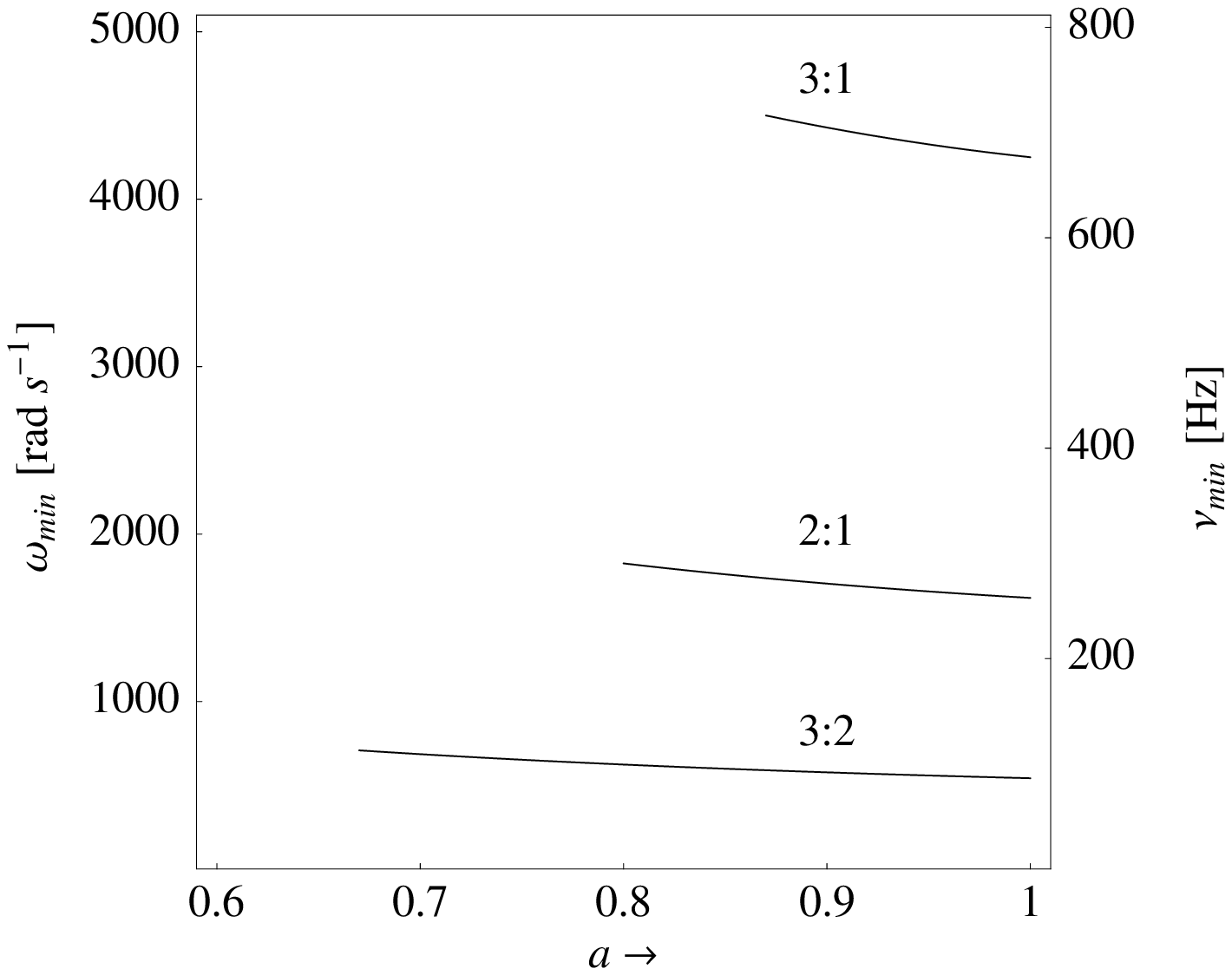, width=20pc}
  \caption{
    Persistent pattern frequencies of three lowest-order transverse
    modes (2:1, 3:1, and 3:2) for a $10M_{\sun}$-mass black hole as a
    function of black hole spin parameter $a$.}
  \label{fig:wa:perp}
\end{figure}

In the last section, we have shown that an accretion disk around a
compact object can support persistent rotating patterns due to the
collective excitations of particles in the disk. Their existence depends
on general relativity effects and is sensitive to the parameters of the
central body but not to the accretion rate. All the frequencies in the
problem scale inversely proportionally to the central body mass. In
addition, persistent pattern frequencies depend on the spin parameter.
Figure~\ref{fig:wa:rad} shows the rotation frequencies of the three
lowest-order persistent patterns (2:1, 3:1, and 3:2) for a
$10M_{\sun}$-mass black hole as the spin is varied. This dependence in
principle could be used to measure the mass and spin of the central
object, provided that the frequencies of two distinct modes are observed
and identified correctly. One should note, though, that for multi-armed
patterns (for example the 3:2 pattern which has two arms) modulation
frequency could be a multiple of the rotation frequency.

Several different persistent patterns could coexist in the accretion
disk; however, it is likely that the lowest-order ones are strongly
selected based on geometrical considerations. Surface density modulation
of the 2:1 pattern is second-order in particle displacement
$\varepsilon$, while the 3:1 and 3:2 patterns are third-order. The
cancellation of lower-order terms is directly caused by the
multiple-fold geometry of the orbits with $n>1$. Being higher-order
could explain why these persistent patterns are not apparent in the
linear perturbation analysis of \citet{2001ApJ...559L..25W}. Despite
second-order scaling, the 2:1 pattern in Figure~\ref{fig:patterns}
(lower right) shows $1\%$ amplitude of the surface density modulation
for moderate displacement values ($\varepsilon/r_* \approx 0.022$). The
radial extent of the 2:1 pattern ($\Delta r \sim 4$) also appears to be
wider than that of a fundamental $g$-mode (the width of which is
proportional to $c_s^{1/2}$ and is estimated as $\Delta r \sim 1$ by
\citet{2001ApJ...559L..25W}).

In this paper, we focused on kinematics and neglected particle
interactions and the hydrodynamics of the disk. The extent to which this
approximation is justified should be further investigated. Of critical
importance for the model is understanding the excitation mechanism. It
is possible for the spiral waves to be driven from the outer edge of the
disk \citep{1987A&A...184..173S,1987MNRAS.229..517S}; however, whether
that is sufficient to cause persistent patterns to appear remains to be
seen. Both questions could be answered by turning to numerical
simulations of the accretion disk hydrodynamics. However, that is a much
more complicated problem, and we feel that it is beyond the scope of
this paper, the intent of which is merely to point out the existence of
new characteristic frequencies in the disk.

It is plausible that the characteristic frequency of the collective
motion will manifest itself in X-ray luminosity variation, but the exact
mechanism responsible for the modulation is not clear to us. Density
pattern in the accretion disk need not be a direct cause. Particles
weaving in and out on self-intersecting orbits could lead to efficient
gas heating, possibly due to shock formation, and create a temperature
pattern in the accretion disk (in a sense, an extensive ``hot spot'').
The picture of temperature modulation of the disk causing X-ray
luminosity variations is not entirely satisfactory as quasi-periodic
oscillations are seen primarily in the hard non-thermal component of the
emission \citep{2003astro.ph..6213M}. That could indicate that the
quasi-periodic emission is coming from a coronal region rather than from
a disk \citep{2004ApJ...612..988T}. It is possible that the transfer
mechanism might involve a magnetic field threading the disk (Blandford,
unpublished).

So far we have been talking about patterns arising from radial
oscillations. It is worth mentioning that a similar thing could happen for
transverse oscillations as well. The precession frequency $\omega_\perp
= \Omega -\frac{n}{m}\,\Omega_\perp$ also has a minimum if $n>m$.
However, the minimum lies inside an innermost stable circular orbit
unless the black hole is spinning rapidly ($a>0.8$ for 2:1 orbit). The
frequencies of the three lowest-order transverse modes are shown in
Figure~\ref{fig:wa:perp}. Transverse particle excitations would lead to
a corrugated accretion disk rather than a surface density pattern.

To summarize, we have found a set of new characteristic frequencies
associated with accretion disks around compact objects. Although many
questions remain, it might be interesting to pursue this idea further
and see if it could lead to a model of quasi-periodic oscillations in
X-ray binaries. In particular, the numerical values of our characteristic
frequencies and their independence of the accretion rate suggest
an application to high-frequency QPOs in black hole binaries.

\section*{Acknowledgments}

We would like to thank Andrei Beloborodov, Roger Blandford, Steven
Fuerst, W{\l}odek Klu\'zniak, and Robert Wagoner for helpful
discussions. AF is supported in part by the Stanford Institute for
Theoretical Physics.

\label{lastpage}
\end{document}